\documentclass[twocolumn,showpacs,10pt,showkeys,amsmath,amssymb]{revtex4-1}

\usepackage{graphicx}
\usepackage{dcolumn}
\usepackage{bm}
\usepackage{verbatim}
\usepackage[version=3]{mhchem}
\usepackage{subfigure}
\usepackage{natbib}
\DeclareMathOperator{\arcsinh}{arcsinh}

\begin{document}

\title{Pt-based nanowire networks with enhanced oxygen-reduction activity}

  
\author{Henning Galinski}
\altaffiliation{Current address: Harvard School of Engineering and Applied Sciences, Harvard University, Cambridge, MA 02138, USA}
\email{hgalinski@seas.harvard.edu}
\affiliation{Nonmetallic Inorganic Materials, ETH Zurich, Zurich, Switzerland}

\author{Thomas Ryll}
\affiliation{Nonmetallic Inorganic Materials, ETH Zurich, Zurich, Switzerland}
\author{Yang Lin}
\affiliation{Nonmetallic Inorganic Materials, ETH Zurich, Zurich, Switzerland}
\author{Barbara Scherrer}
\affiliation{Nonmetallic Inorganic Materials, ETH Zurich, Zurich, Switzerland}
\author{Anna Evans}
\affiliation{Nonmetallic Inorganic Materials, ETH Zurich, Zurich, Switzerland}
\author{ Max D\"obeli}
\affiliation{Ion Beam Physics, ETH Zurich, Zurich, Switzerland}
\author{Ludwig J. Gauckler}
\affiliation{Nonmetallic Inorganic Materials, ETH Zurich, Zurich, Switzerland}

\begin{abstract}
Pt-Al and Pt-Y-Al thin film electrodes on yttria-stabilised zirconia electrolytes were prepared by dealloying of co-sputtered Pt-Al or Pt-Y-Al films. The selective dissolution of Al from the Pt-alloy compound causes the formation of a highly porous nanowire network with a mean branch thickness below $25$~nm and a pore intercept length below $35$~nm. The oxygen reduction capability of the resulting electrodes was analysed in a micro-solid oxide fuel cell setup at elevated temperatures ($598-873$K). Here, we demonstrate that these nanoporous thin films excel "state-of-the-art" fuel cell electrodes in terms of catalytic activity and thermal stability. The nanoporous Pt electrodes exhibit exchange current densities that are up to 13 times higher than conventional Pt electrodes, measured at $648$~K. It is shown that the enhanced catalytic activity of these Pt electrodes is achieved through the engineering of the materials d-bands due to the addition of yttrium as ternary constituent.\\

\end{abstract}
\maketitle

\section*{Introduction}

Pt has outstanding catalytic properties, and it is used in numerous modern technologies, ranging from catalytic converters~\cite{Farkas1,Miller1} in motor vehicles, chemical reactors and electrochemical devices, such as sensors, batteries and fuel cells~\cite{Scherrer1,Stamenkovic1,Steele1}. One important goal of the present research efforts is to design new materials that allow for a reduction of the Pt content without sacrificing the activity, selectivity and stability that Pt guarantees for. A common approach in catalyst design is the so-called d-band engineering~\cite{Greeley2,Viswan1,Johannesson1} where Pt is experimentally or computationally~\cite{Norskov1} alloyed by another catalytic active element such as Rh~\cite{Greeley1}, Ru~\cite{Jiang1}, Ni~\cite{Stamenkovic2,Snyder1}, Ir~\cite{Greeley1}, Y~\cite{Greeley1}, Pd~\cite{Tong1,Stamenkovic2} or Co~\cite{Greeley1,Stamenkovic1,Oezaslan1,Snyder2}. The electronic structure, the d-bands in particular, is altered and with it the adsorption energy of any adsorbing species on the Pt-alloy surface. The alloy material can lower the cost and trigger the selectivity of the Pt-alloy surface in such a way, that only one type of molecules is adsorbed and catalysed.\\ 
In a solid oxide fuel cell, the Pt-alloy catalyst needs to exhibit long-term stability in hydrogen/hydrocarbon and oxygen-rich environments at temperatures up to $873$~K ($600^{\circ}$C), as well. The energy conversion in the cell is achieved by two separated reactions: the hydrogen oxidation reaction (HOR) at the anode and the oxygen reduction reaction (ORR) at the cathode. The ORR is typically the rate-limiting step of the micro-SOFC at low temperatures~\cite{Debe1} and involves multiple reaction steps~\cite{Hoerlein1}, namely oxygen adsorption, surface diffusion of intermediates, charge transfer and oxygen incorporation into the electrolyte, whose relative speed and spatial arrangement determine the overall reaction mechanism~\cite{Norskov2,Debe2}. Insight in the processes that limit the electrode performance is gained by comparing the obtained experimental current-voltage curves to kinetic models that address the electrochemical processes in the fuel cell.
\begin{figure*}[t!]
  \begin{center}
	  \subfigure[]{\label{fig:nature1-a}\includegraphics[scale=0.65]{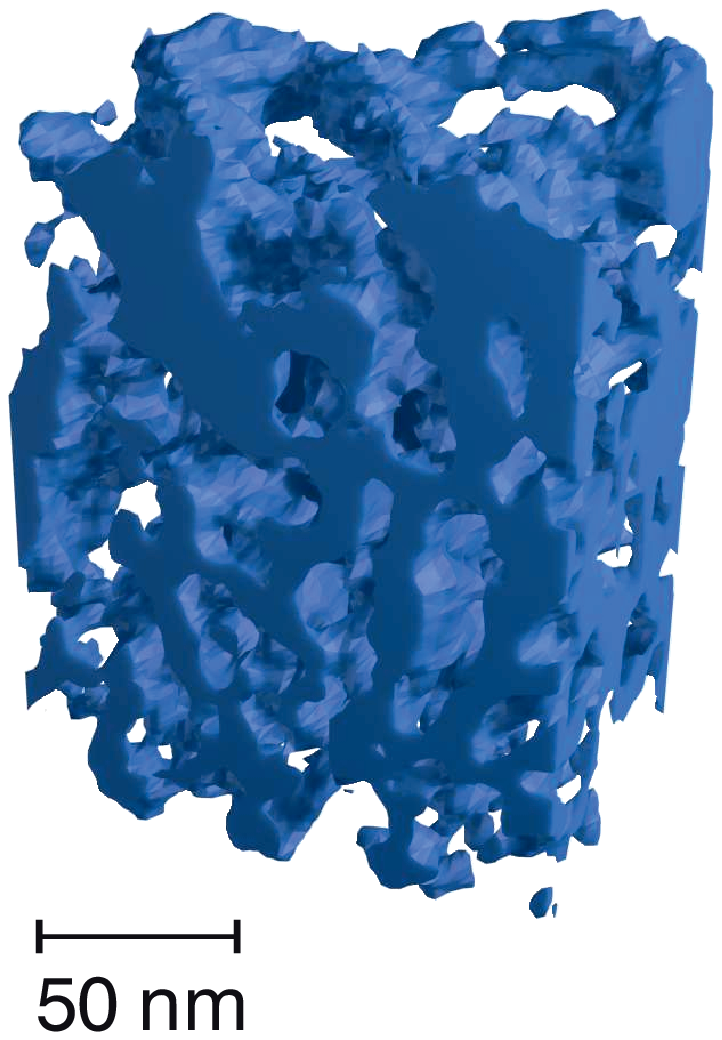}}
		\hspace{1.0cm}
		  \subfigure[]{\label{fig:nature1-b}\includegraphics[scale=1.2]{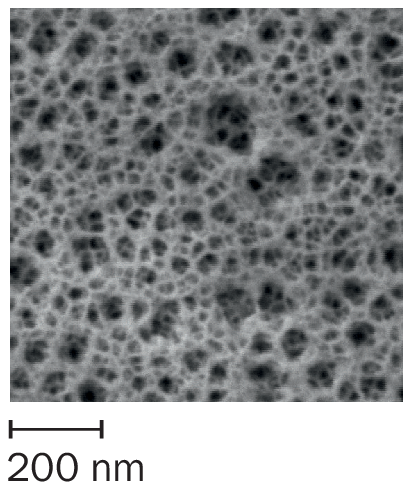}}
				\\
		\subfigure[]{\label{fig:nature1-e}\includegraphics[scale=0.525]{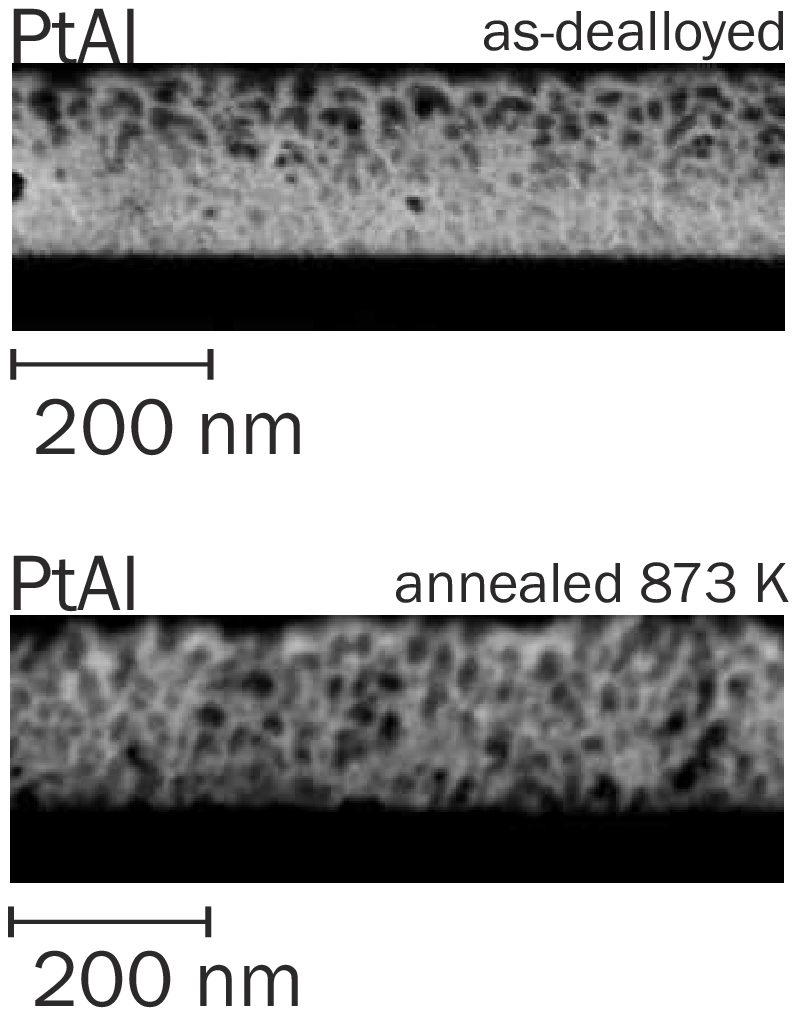}}
		\hspace{0.75cm}
    \subfigure[]{\label{fig:nature1-c}\includegraphics[scale=0.52]{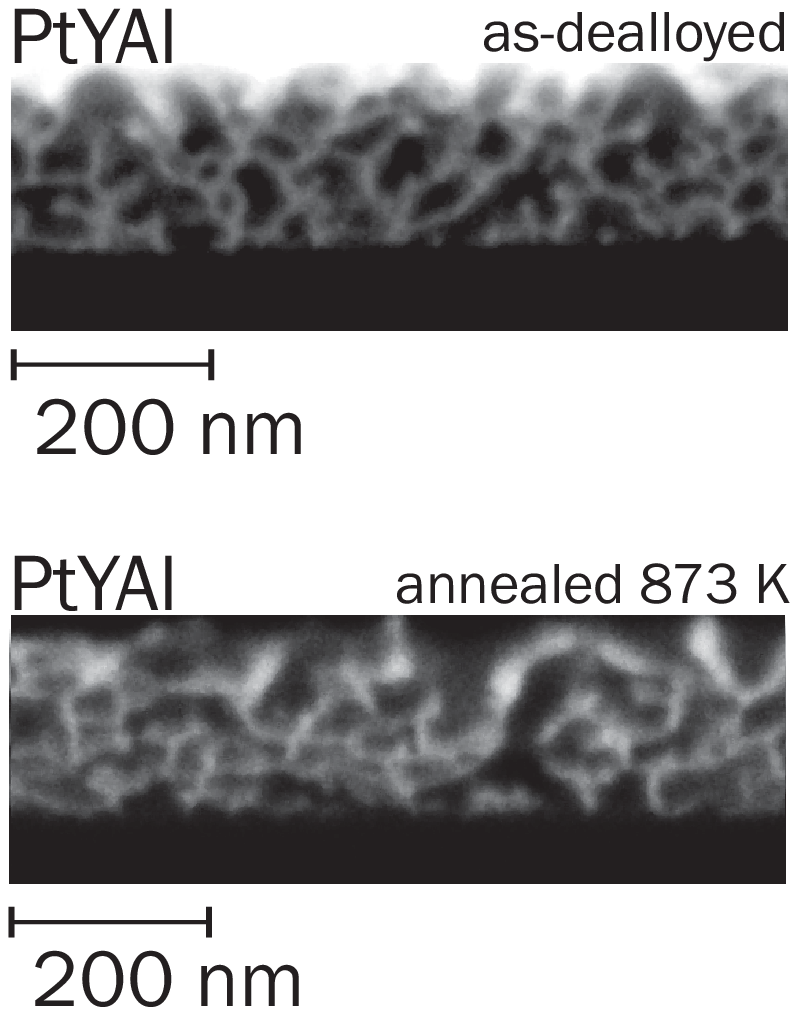}}
		\hspace{0.75cm}
		\subfigure[]{\label{fig:nature1-d}\includegraphics[scale=0.275]{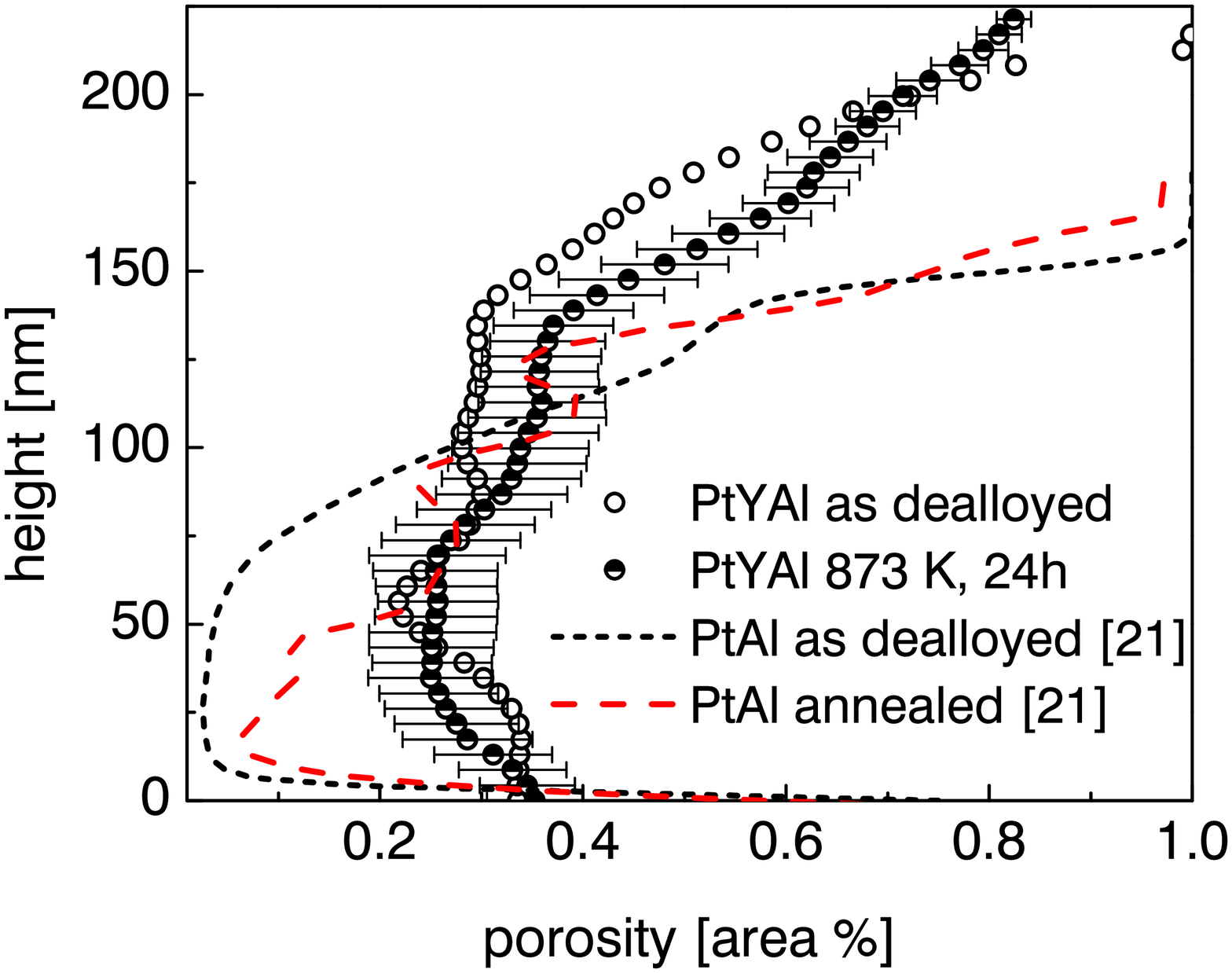}}
  \end{center}
  \caption{Nanowire Networks.~\subref{fig:nature1-a}, Three dimensional (3D) reconstructed FIB thin-film tomography of a Pt$_{.70}$Al$_{.30}$ nanowire network on a YSZ single crystal.~\subref{fig:nature1-b}, Typical scanning electron microscopy top-view image of a nanoporous Pt$_{.70}$Al$_{.30}$ nanowire network. The nanowires are interconnected and are characterised by a mean length of $30$~nm and a mean diameter of $12$~nm. \subref{fig:nature1-e}-\subref{fig:nature1-c}, FIB-polished cross-sections of Pt$_{.70}$Al$_{.30}$ and Pt$_{.60}$Y$_{.26}$Al$_{.14}$ nanowire networks on a YSZ single crystal before and after annealing at $873$~K, $24$~h. \subref{fig:nature1-d}, Porosity before and after annealing of a Pt$_{.60}$Y$_{.26}$Al$_{.14}$ film plotted as a function of the film height perpendicular to the film/substrate interface and compared to the porosity of Pt$_{.72}$Al$_{.28}$ films from Ref.~\cite{Ryll1}.}
\label{fig:nature1}
\end{figure*}
\begin{figure*}[t!]
  \begin{center}
    \subfigure[]{\label{fig:nature3-a}\includegraphics[width=.85\textwidth]{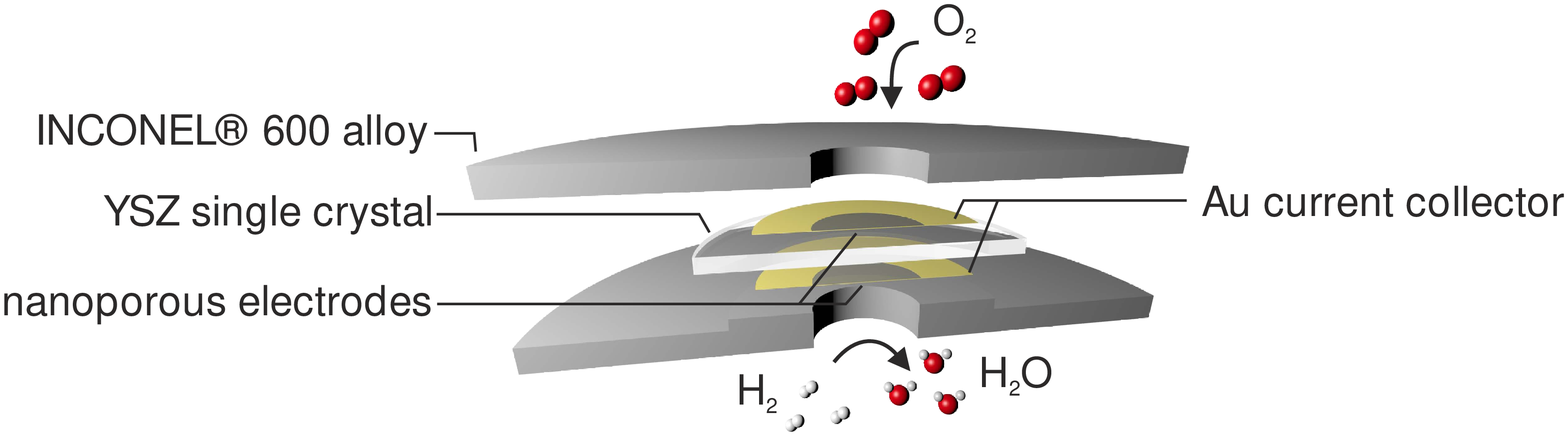}}
		\\
   	\subfigure[]{\label{fig:nature3-b}\includegraphics[width=.5\textwidth]{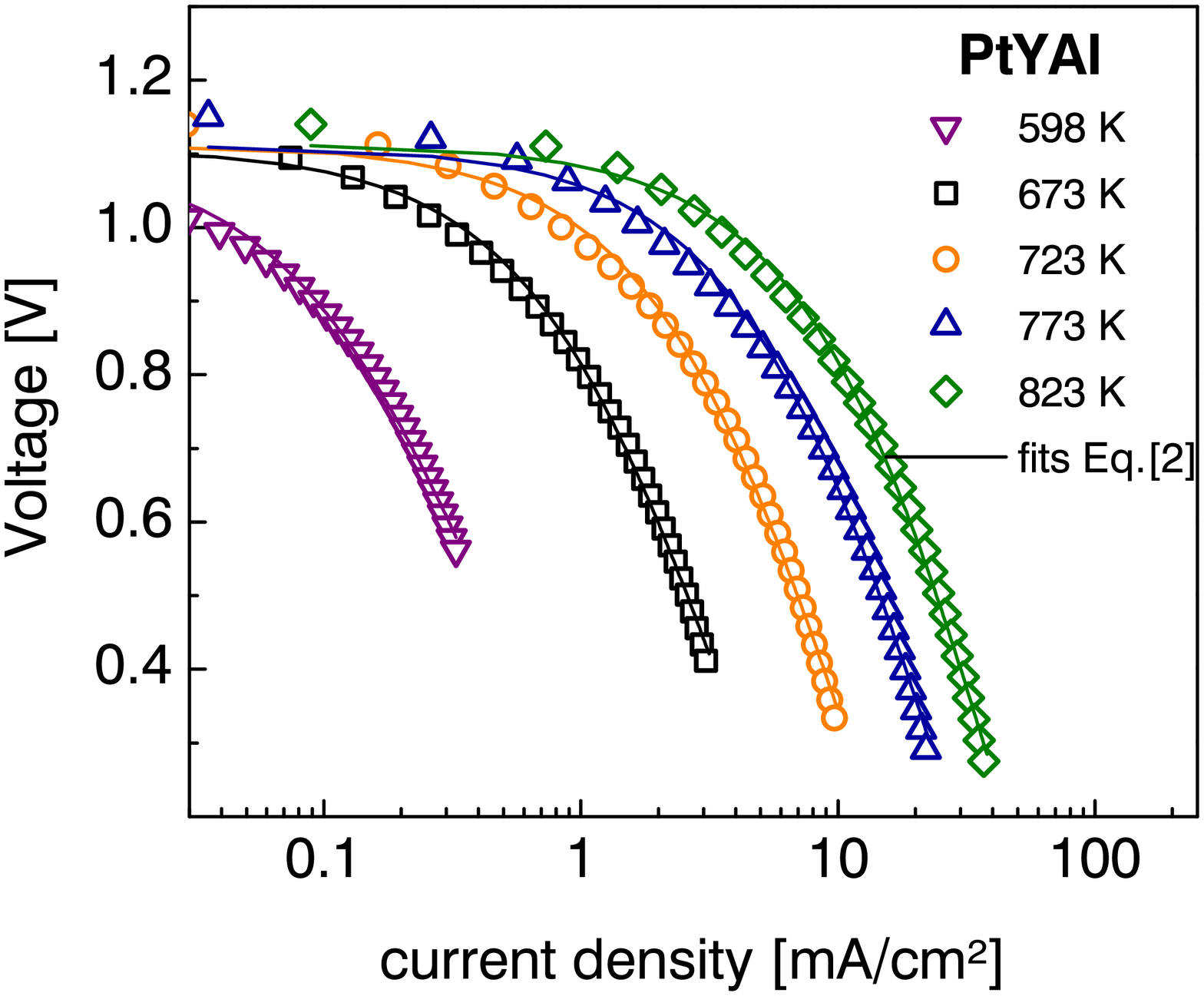}}
		\subfigure[]{\label{fig:nature3-c}\raisebox{3.5mm}{\includegraphics[width=.49\textwidth]{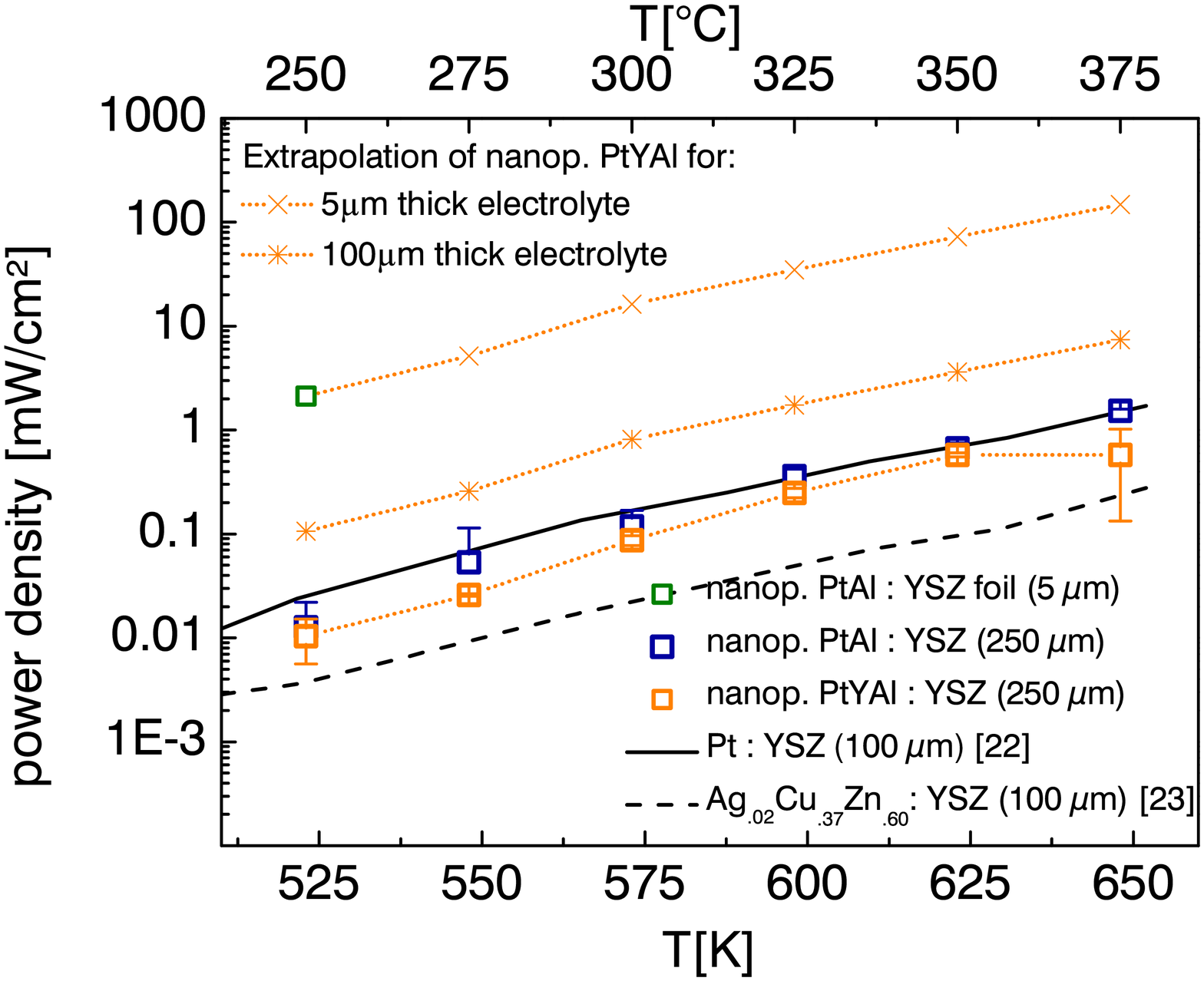}}}
  \end{center}
  \caption{Fuel cell setup and device performance.~\subref{fig:nature3-a}, Schematic illustration of the experimental fuel cell setup~\subref{fig:nature3-b}. Current-voltage characteristics of the Pt$_{.60}$Y$_{.26}$Al$_{.14}$ fuel cells fitted by Eq.~\ref{eq:HCLimit} (lines) at five different temperatures. \subref{fig:nature3-c}, Measured peak power density $\rho$ as function of the operating temperature compared to data from literature~\cite{Chao1,Holme1}. The orange dotted lines with symbols show an extrapolation of the power density (Pt-Y-Al electrodes) for two different electrolyte thicknesses (100$\mu$m,5$\mu$m). The error bars represent the standard deviation between different cells.} 
  \label{fig:nature3}
\end{figure*}
The chemical conversion of energy in the cell and result in a cell voltage $V$ that is lower than its ideal potential $V_0$. The electrode-related losses in cell voltage as a function of the current density $i$ can be described by the Tafel equation~\cite{FuelCells1}, 
\begin{equation}
V = V_0 - \frac{R T}{\alpha n F} \log{\frac{i}{i^{\star}}}.
\label{eq:TafelExpression}
\end{equation}
The exchange current density $i^{\star}$ relates to processes at both anode and cathode. $\alpha$ is the electron transfer coefficient. $R,T,n,F$ have their usual meaning. In order to separate the electrochemical reactions at the anode and cathode, the cell voltage can be expressed by a Butler-Volmer model~\cite{Biesheuvel1}, it reads
\begin{equation}
V = V_0 - \left\{ \frac{2}{f} \arcsinh \frac{i}{2 i^{\star}_\text{anode}} + \frac{2}{f} \arcsinh \frac{i}{2 i^{\star}_\text{cathode}} \right\}.
\label{eq:HCLimit}
\end{equation}
Thereby $f$ equals $F/R T$ and $i_\text{anode}$ and $i_\text{cathode}$ are the electrode specific exchange current densities. The exchange current density reflects the catalytic activity of the electrode~\cite{Debe2} and is well suited to compare different nanostructures and electrode materials even when measured with different electrolytes.\\
Here, nanowire networks assembled from co-sputtered Pt-Al and Pt-Y-Al thin films are presented as an ideal candidate for high-performance fuel cell electrodes. 
Briefly, two different electrodes have been tested in a symmetric electrolyte-supported fuel cell setup (Fig.~\ref{fig:nature3-a}): Nanoporous Pt-Al thin film electrodes and nanoporous Pt-Y-Al thin film electrodes. Both electrodes were deposited at room temperature by magnetron co-sputtering onto single crystalline yttrium-stabilised zirconia (\ce{YSZ}) substrates and in the case of the Pt-Al electrodes also on $5$~$\mu$m thin ceramic YSZ foils~\cite{Bonderer1, Evans1}. A detailed description is given in the Supplemental Material~\cite{Supplemental}. The assembly of the nanowire networks is achieved by the selective dissolution~\cite{Galinski1,Erlebacher1,Supansomboon1} of the less noble constituent (Al) of the Pt-alloy thin film by immersing the coated substrates for up to $60$~s in a $4 M$ aqueous solution of \ce{NaOH}. This process is a simple, fast and cost-effective route to form scalable metal-based nanowire networks.\\ The selective dissolution of the Al causes a re-assembly of the metal matrix which is characterised by a blood-vessel like structure composed of connected nanowires with a length up to $50$~nm and typical diameters of $\approx 15$~nm. In thin films, this pattern formation is mainly caused by a linearly propagating diffusion front, i.e. the liquid/solid interface, travelling through the film at a constant speed~\cite{Galinski1}.

\section*{Results and Discussion}

\subsection*{Morphology and structure}

In Figure~\ref{fig:nature1-a} and ~\ref{fig:nature1-b}, the morphology of a dealloyed Pt$_{.70}$Al$_{.30}$ nanowire network obtained thin-film tomography and by scanning electron microscopy is shown. The re-assembly of the metal matrix due to the dissolution of Al results in a network structure with more than $1800$ nanowires per $\mu$m$^3$. The nanowires have a mean length of $30$~nm and a mean diameter of $12$~nm. 
These electrodes have already a high surface area for fuel absorption and are good electrical conductors but unfortunately undergo both structural and electrochemical degradation that is attributed to the oxidation of residual Al in the nanowires~\cite{Ryll1}. In order to overcome these problems, yttrium has been added as ternary constituent. The addition of yttrium to an alloy is known to increase both the thermal stability and the resistance to high-temperature oxidation~\cite{Johannesson1}. Moreover, it has been shown by Greeley\textit{ et al.}, that the oxygen-reduction activity of bulk \ce{Pt3Y} electrodes is up to $10$ times higher than the activity of pure Pt~\cite{Greeley1}. In Fig~\ref{fig:nature1-c} the cross-sectional microstructures of a dealloyed Pt-Y-Al thin film is shown prior and after annealing at $873$~K for $24$~h. Compared to the Pt-Al thin films (Fig.~\ref{fig:nature1-e}), the addition of yttrium changes the morphology of the nanowire network significantly. These changes manifest in a decrease of the nanowire density to $400$ nanowires per $\mu$m$^3$, which also decreases the surface area of the electrode. Furthermore, the mean branch diameter increases to $25$~nm, alongside with an enhancement of the mean pore intercept length from $10$~nm to $35$~nm. Despite the decrease in surface area, the addition of yttrium has improved the thermal stability significantly. This is illustrated in Fig.~\ref{fig:nature1-c} presenting the nanostructure of as-dealloyed and annealed Pt-Y-Al nanowire networks. The effect becomes even more evident when the porosity determined by means of multiple cross-sections of as-dealloyed and annealed Pt-Y-Al films is plotted as a function of film height perpendicular to the film/substrate interface (see Fig~\ref{fig:nature1-d}). In contrast to Pt-Al networks, the structural integrity of Pt-Y-Al is conserved after annealing the films at $873$~K for $24$~h, cf. Fig~\ref{fig:nature1-d}. Thus, the gain in thermal stability could be attributed to the compositional changes in the nanowire network.
\begin{figure}[h!]
  \begin{center}
   \includegraphics[width=.49\textwidth]{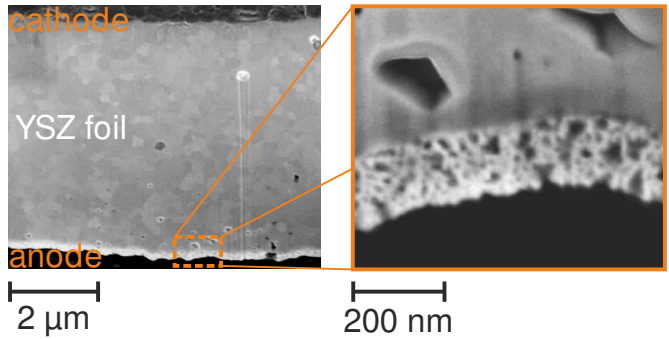}
  \end{center}
  \caption{FIB-polished cross-sections of a symmetric fuel cell with Pt$_{.70}$Al$_{.30}$ thin films as electrodes deposited on a $5$~$\mu$m YSZ foil with a mean grain size of $250$~nm. The detail of the picture shows the electrode's nanostructure.} 
  \label{fig:nature5}
\end{figure}
Indeed, this has been confirmed by Rutherford backscattering spectroscopy showing that the added yttrium replaces the aluminium in the nanowire network and causes a change in composition from Pt$_{.70}$Al$_{.30}$ to Pt$_{.60}$Y$_{.26}$Al$_{.14}$. 	In conjunction with the reduced Pt content due to the addition of yttrium, the Pt-load is reduced from $166$~$\mu$g/cm$^2$ to $142$~$\mu$g/cm$^2$, assuming a $300$~nm thick film and a mean porosity of $0.4$.
The found composition of Pt and Y are in excellent agreement with the nominal composition of the \ce{Pt3Y} intermetallic phase of the Pt-Y system~\cite{Palenzona1}. It is noteworthy that density functional theory (DFT) calculations predicted the \ce{Pt3Y} alloy to be one of the most stable of all fcc-alloys, where all bonding states are occupied and the anti-bonding states are empty~\cite{Johannesson1}.\\

\subsection*{Electrochemical activity}

Fuel cells were constructed using both the Pt-Al and Pt-Y-Al electrodes with an active surface area of $26.6$~mm$^2$ on $250$~$\mu$m thick YSZ single crystal electrolytes. A sketch of the fuel cell design is shown in Fig.~\ref{fig:nature3-a}. The cells were tested with air on the cathode side and a \ce{H2}/\ce{N2} mixture on the anode side at operation temperatures between $523-873$~K. For each temperature, current-voltage curves of the fuel cell were measured. A typical set of current-voltage curves as function of temperature for a cell with Pt-Y-Al electrodes is given in Fig.~\ref{fig:nature3-b}. The devices are characterised by an open circuit voltage of $V_\text{OCV} = 1.10-1.15$~V and a maximal power density of $P_\text{max} = 4.6 $~mW/cm$^2$ at $873$~K. The current-voltage characteristics for fuel cells with Pt-Al electrodes are included in the Supplemental Material~\cite{Supplemental}.
The corresponding maximal power density compared to data from literature is shown in Figure~\ref{fig:nature3-c}. 
The power density of our fuel cells on $250$~$\mu$m thick YSZ electrolytes falls together with the power densities measured for standard Pt electrodes on $100$~$\mu$m YSZ~\cite{Chao1}. As the overall fuel cell's power density scales linearly with the thickness of a purely ionic conducting electrolyte~\cite{Kerman1,Biesheuvel1,An1}, it can be concluded that the electrode performance of herein analysed Pt-based nanowire networks is roughly a factor $3$ higher than the conventional nanoporous Pt electrodes~\cite{Chao1} and non-precious electrodes~\cite{Holme1}. As proof that the electrolyte thickness is indeed the limiting factor for the cell-performance, symmetric fuel cells with Pt-Al electrodes and a $5$~$\mu$m thin ceramic YSZ foil~\cite{Bonderer1, Evans1} as electrolyte were fabricated as well. The FIB cross-sectional image of such a fuel cell shown in Fig.~\ref{fig:nature5} reveals a dense electrolyte with a homogeneous microstructure and a mean grain size of about $250$ nm. Although the surface of the electrolyte is rough compared to the YSZ single crystals, the Pt-Al electrode shown in the enlarged part of Fig.~\ref{fig:nature5} is percolating and the nanowire network is crack-free. Alongside with the power density of a theoretical fuel cell with an electrolyte thickness of $5$ and $100$~$\mu$m, the power density for a fuel cell with a thin YSZ foil as electrolyte and nanoporous Pt-Al electrodes is shown in Fig.~\ref{fig:nature3-c}. The found maximal power density of $2.1$~mW/cm$^2$ at $523$~K for the Pt-Al electrodes on the YSZ foil is in excellent agreement with the calculated value (see Supplemental Material~\cite{Supplemental} for additional details).\\ 
\begin{figure}[t!]
  \begin{center}
    \subfigure[]{\label{fig:nature4-a}\includegraphics[scale=0.3]{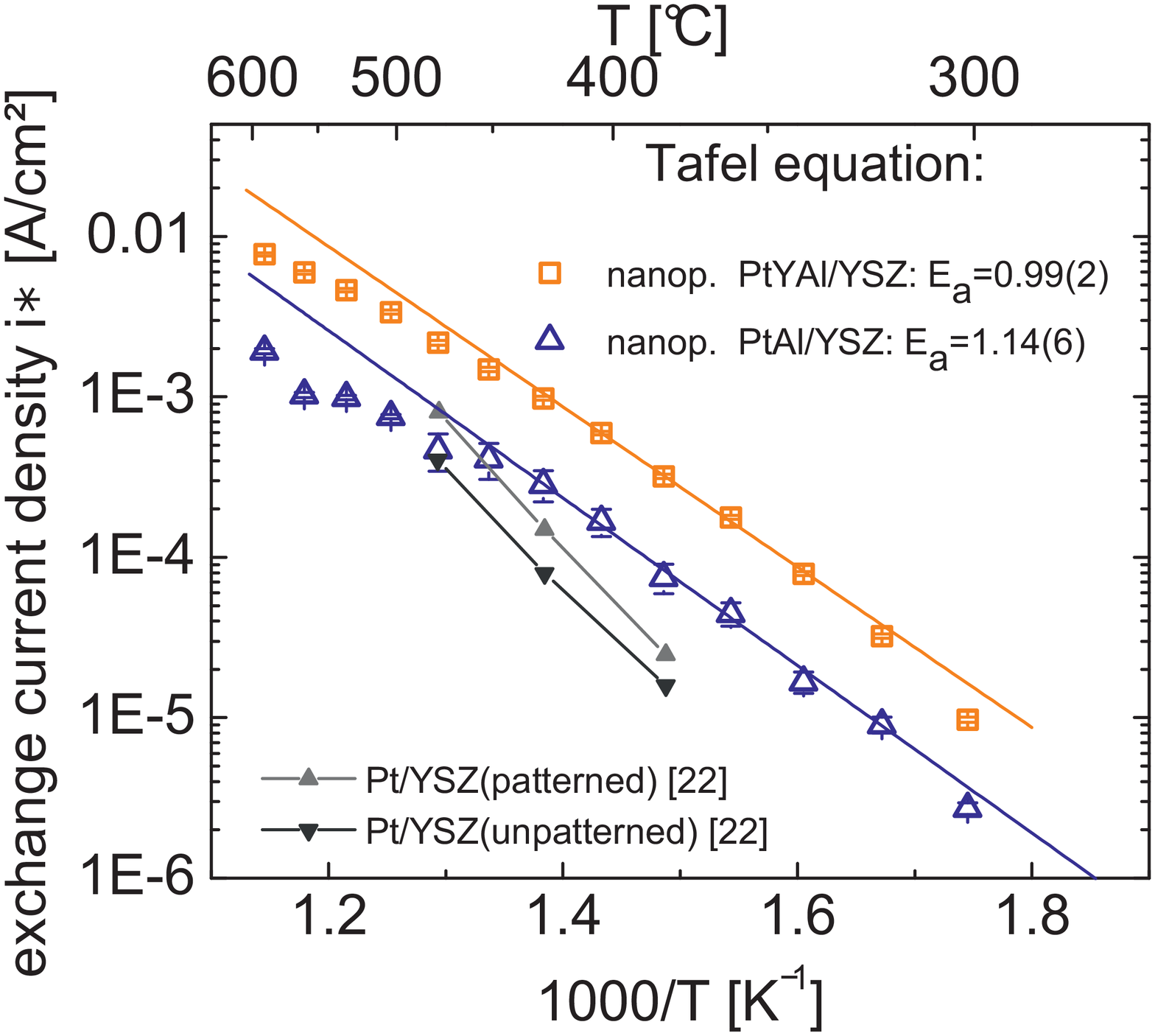}}
    \\
    \subfigure[]{\label{fig:nature4-b}\includegraphics[scale=0.32]{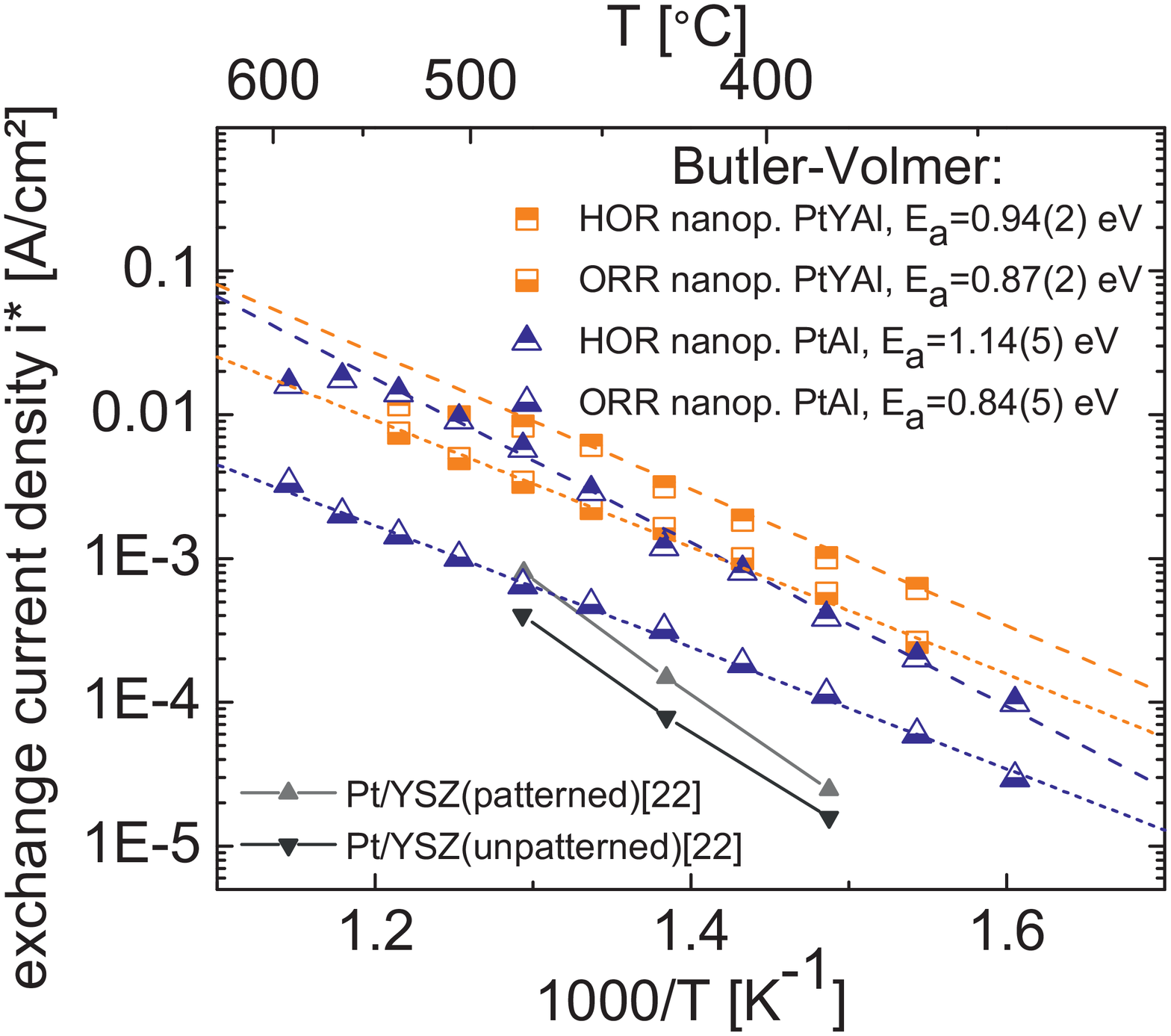}}
  \end{center}
  \caption{Exchange current density and thermal activation. \subref{fig:nature4-a} Exchange current density $i^{\ast}$ determined by fitting the Tafel equation (Eq.\ref{eq:TafelExpression}) to the cells' I-V curves. \subref{fig:nature4-b} Exchange current density $i^{\ast}$ for the ORR and HOR reactions obtained from fits using the Butler-Volmer model (Eq.\ref{eq:HCLimit}).} 
  \label{fig:nature3}
\end{figure}
In order to assess the electrode specific kinetics, the I-V curves have been fitted by the Tafel equation (Eq.~\ref{eq:TafelExpression}) and the Butler-Volmer model (Eq.~\ref{eq:HCLimit}). Figure~\ref{fig:nature4-a} shows an Arrhenius plot of the estimated exchange current densities $i^\star$ obtained for the overall kinetics by the Tafel equation. Compared with conventional Pt electrodes the exchange current densities for our best electrodes are a factor $3$ (Pt-Al) and $13$ (Pt-Y-Al) higher at $648$~K. At higher temperatures, the exchange current density between the conventional electrodes~\cite{Chao1} and the Pt-Y-Al electrodes differs by only a factor of $3$. Nonetheless, this is still remarkable in terms of reactivity as the Pt content is significantly reduced from $166$~$\mu$g/cm$^2$ to $142$~$\mu$g/cm$^2$ while the stability of the electrodes has been increased.\\
In order to gain more insight on the specific kinetics on the cathode side (ORR) and the anode side (HOR), we employed the Butler-Volmer model (Eq.~\ref{eq:HCLimit}) and fitted the measured I-V curves. In Fig.~\ref{fig:nature3-a}, the I-V curves of a cell with Pt-Y-Al electrodes at different operating temperatures is depicted together with the fitting curves from the model. The fitting curves generated from the employed Butler-Volmer model are in close agreement with the measured I-V curves for all operating temperatures. The measured I-V curves including the fitted model curves for a cell with Pt-Al electrodes are available in the Supplemental Material~\cite{Supplemental}.\\
The Butler-Volmer model allows to separate the kinetics at the two electrodes and to determine their specific exchange current density $i^{\star}_{\text{anode,cathode}}$, see Fig.~\ref{fig:nature4-b}. The measured HOR for the Pt-Al and the Pt-Y-Al electrode are considerably different not only in the absolute value of the exchange current density but also in the overall thermal activation, that is reduced from $E_\text{a,Pt-Al}=1.14$~eV to $E_\text{a,Pt-Y-Al}=0.94$~eV as consequence of the yttrium addition. On the cathode side, solely the absolute value of the exchange current density, thus the oxygen reactivity is positively affected. This considerable enhancement in ORR and HOR activity for PtYAl can be attributed to the reduced oxygen adsorption energy and hydrogen adsorption energy on \ce{Pt3Y} in respect to pure Pt~\cite{Greeley1}. The enhanced oxygen reduction activity of almost one order of magnitude agrees well with literature data from Ref.~\cite{Greeley1}, whereby the enhanced activity can be understood in terms of the d-band model, as the addition of the early transition metal yttrium lowers the d-band centre with respect to the Fermi level, which decreases the binding energy between the catalyst and \ce{O2} or \ce{H2} respectively and therefore the overpotential of the ORR and HOR~\cite{Greeley1}.\\

\section*{Conclusion}

Nanoporous binary and ternary Pt-alloy electrodes are an exciting and promising candidate as electrodes for fuel cells and other electrochemical devices. We have analysed the electrochemical activity and thermal stability of Pt-Al und Pt-Y-Al nanowire networks operating as electrodes in a solid oxid fuel cell setup. It has been demonstrated that these electrodes show exceptional catalytic activity that excells conventionally designed catalysts by up to a factor of $13$. In the case of the analysed \ce{Pt3Y} electrodes, we demonstrate that the addition of the catalytic active element Y impairs the catalytic activity of the nanowire network through d-band engineering. This allows for a reduction of the Pt-load from $166$~$\mu$g/cm$^2$ to $142$~$\mu$g/cm$^2$ without sacrificing the electrochemical activity of the electrodes.\\
Furthermore, it is shown that the \ce{Pt3Y} electrodes feature an enhanced thermal stability and withstand temperatures as high as $823$~K in oxidising atmosphere for $24$~h. In view of other electrode applications these result are beneficial for the improvement of catalytic converters in motor vehicles, as the alloys (Pt-Rh, Pt-Pd, Pd-Rh) which are commonly used, suffer from significant degradation at this temperature~\cite{Tong1}. The observed scaling of both the catalytic and thermal properties of the Pt and \ce{Pt3Y} nanowire electrodes are in excellent agreement with findings for bulk-catalysts of the same type~\cite{Greeley1}.\\
Further studies will focus on the impact of geometrical confinement on the catalytic properties at the nanoscale by examining the catalytic activity of nanowire networks with a mean branch diameter below $15$~nm.

\begin{acknowledgements}
H. G. and B.S. gratefully acknowledge financial support from the \textit{Size Matters!} project, Switzerland.\\
The authors would like to thank the EMEZ (Electron Microscopy Center, ETH Zurich) and the FIRST cleanroom team for their support. H.G. thanks for P. Gasser for outstanding FIB maintenance and I. Schenker for peripatetic discussions\\
The authors declare that they have no competing financial interests.\\
\end{acknowledgements}

\bibliographystyle{apsrev4-1}
\bibliography{bibfile} 

\end{document}